\documentclass[apj]{emulateapj}
\usepackage{apjfonts}
\usepackage{amsbsy}

\def\hmpc{$h^{-1}$Mpc}
\def\hkpc{$h^{-1}$kpc}

\def\mstar{$M_\ast$}

\def\hmsol{$h^{-1}$M$_\odot$}

\def\om{\Omega_m}

\def\s8{\sigma_8}

\def\plin{P_{\rm lin}(k)}

\def\lcdm{$\Lambda$CDM}

\def\x2{$\chi^2$}
\def\hmsol{$h^{-1}\,$M$_\odot$}

\def\NNm1{\langle N(N-1) \rangle}

\def\dc{\delta_c}

\def\m_star{M_\ast}

\def\plin{P_{\rm lin}(k)}

\def\lcdm{$\Lambda$CDM}

\def\om{\Omega_m}

\def\s8{\sigma_8}

\def\hmpc{$h^{-1}\,$Mpc}
\def\hkpc{$h^{-1}\,$kpc}

\def\x2{$\chi^2$}
\def\hmsol{$h^{-1}\,$M$_\odot$}

\def\plin{P_{\rm lin}(k)}

\def\mstar{M_\ast}

\def\NNm1{\langle N(N-1) \rangle}

\def\dc{\delta_c}

\def\dc{\delta_c}

\def\mstar{M_\ast}

\def\p0{P_0(r)}

\def\s{\nu}

\def\D{\Delta}
\def\rd{R_{\Delta}}
\def\xdof{\chi^2/\nu}

\def\bhm{b_{hm}}
\def\chinu{\chi^2_\nu}

\newcommand{\deltac}{\delta_{c}}
\newcommand{\deltahL}{\delta_{\mathrm{h}}^{\mathrm{L}}}

\begin{document}

\title{The Large Scale Bias of Dark Matter Halos: Numerical Calibration and Model Tests}

\author{
Jeremy L. Tinker\altaffilmark{1}, Brant E. Robertson\altaffilmark{2,10},
Andrey V. Kravtsov\altaffilmark{3,4,5},
Anatoly Klypin\altaffilmark{6},\\
Michael S. Warren\altaffilmark{7},
Gustavo Yepes\altaffilmark{8},
Stefan Gottl{\"o}ber\altaffilmark{9}
}
\altaffiltext{1}{Berkeley Center for Cosmological Physics, University of California, Berkeley}
\altaffiltext{2}{Astronomy Department, California Institute of Technology, MC 249-17, 1200 East California Boulevard, Pasadena, CA 91125}
\altaffiltext{3}{Kavli Institute for Cosmological Physics, The University of Chicago, 5640 S. Ellis Ave., Chicago, IL 60637, USA}
\altaffiltext{4}{Department of Astronomy \& Astrophysics, The University of Chicago, 5640 S. Ellis Ave., Chicago, IL 60637, USA}
\altaffiltext{5}{Enrico Fermi Institute, The University of Chicago, 5640 S. Ellis Ave., Chicago, IL 60637, USA}
\altaffiltext{6}{Department of Astronomy, New Mexico State University}
\altaffiltext{7}{Theoretical Astrophysics, Los Alamos National Labs} 
\altaffiltext{8}{Grupo de Astrofísica, Universidad Autónoma de Madrid}
\altaffiltext{9}{Astrophysikalisches Institut Potsdam, Potsdam, Germany}
\altaffiltext{10}{Hubble Fellow}

%%%%%%%%%%%%%%%%%%%%%%%%%%%%%%%%%%%%%%%%%%%%%%%%%%%%%%%%%%%%%%%%%%%%%%%
% ABSTRACT ABSTRACT ABSTRACT ABSTRACT ABSTRACT ABSTRACT ABSTRACT 
%%%%%%%%%%%%%%%%%%%%%%%%%%%%%%%%%%%%%%%%%%%%%%%%%%%%%%%%%%%%%%%%%%%%%%%

\begin{abstract}

  We measure the clustering of dark matter halos in a large set of
  collisionless cosmological simulations of the flat \lcdm\
  cosmology. Halos are identified using the spherical overdensity
  algorithm, which finds the mass around isolated peaks in the density
  field such that the mean density is $\D$ times the background. We
  calibrate fitting functions for the large scale bias that are
  adaptable to any value of $\D$ we examine. We find a $\sim 6\%$
  scatter about our best fit bias relation.  Our fitting functions
  couple to the halo mass functions of Tinker et.~al.~(2008) such that
  bias of all dark matter is normalized to unity. We demonstrate that
  the bias of massive, rare halos is higher than that predicted in the
  modified ellipsoidal collapse model of Sheth, Mo, \& Tormen (2001),
  and approaches the predictions of the spherical collapse model for
  the rarest halos. Halo bias results based on friends-of-friends
  halos identified with linking length 0.2 are systematically lower
  than for halos with the canonical $\D=200$ overdensity by $\sim
  10\%$. In contrast to our previous results on the mass function, we
  find that the universal bias function evolves very weakly with
  redshift, if at all. We use our numerical results, both for the mass
  function and the bias relation, to test the peak-background split
  model for halo bias. We find that the peak-background split achieves
  a reasonable agreement with the numerical results, but $\sim 20\%$
  residuals remain, both at high and low masses.

\end{abstract}


%%%%%%%%%%%%%%%%%%%%%%%%%%%%%%%%%%%%%%%%%%%%%%%%%%%%%%%%%%%%%%%%%%%%%%%
% INTRODUCTION INTRODUCTION INTRODUCTION INTRODUCTION INTRODUCTION 
%%%%%%%%%%%%%%%%%%%%%%%%%%%%%%%%%%%%%%%%%%%%%%%%%%%%%%%%%%%%%%%%%%%%%%%

\section{Introduction}

Dark matter halos are biased tracers of the underlying dark matter
distribution. Massive halos form from high-$\sigma$ fluctuations in
the primordial density field, inducing a correlation between halo mass
and clustering amplitude that is steepest for cluster-sized objects
(\citealt{kaiser:84}). Low-mass halos are preferentially found in
regions of the universe with below average density, thus these objects
are anti-biased with respect to the dark matter. The clustering of
galaxies is now understood through the bias of the halos in which they
form (e.g., \citealt{zehavi_etal:05}). Many methods that utilize
galaxy clustering to constrain cosmology require precise knowledge of
halo clustering (e.g., \citealt{vdb_etal:03, tinker_etal:05,
  kev_etal:05, zheng_weinberg:07, yoo_etal:09}). Cosmological
parameters can also be obtained through the abundance of high-mass
halos identified as galaxy clusters. The bias of clusters contains
complementary information to their abundance. Indeed,
``self-calibration'' of cluster surveys is not possible without the
additional information present in clustering data
(\citealt{lima_hu:04, lima_hu:05, majumdar_mohr:04, oguri:09}). The
purpose of this paper is to calibrate a precise, flexible halo bias
function from numerical simulations that is accurate for dwarf
galaxies through galaxy cluster masses.

In \cite{tinker_etal:08_mf} (hereafter T08), we presented a
recalibration of the halo mass function based on a large series of
collisionless N-body simulations. Our results utilized the spherical
overdensity (SO) algorithm for identifying dark matter halos within
simulations (e.g., \citealt{lacey_cole:94}). In this approach, halos
are identified as isolated density peaks, and the mass of a halo is
determined by the overdensity $\D$, defined here as the mean interior
density relative to the background. Simulations of cluster formation
show that the SO-defined halo mass should correlate tightly with
cluster observables, which are usually defined within a spherical
aperture (e.g., \citealt{bialek_etal:01, da_silva_etal:04, nagai:06,
  kravtsov_etal:06}). This expectation is borne out for observables
such as gas mass, core-excised luminosity, integrated SZ flux or its
X-ray analog, $Y_{\rm X}$ (e.g., \citealt{mohr_etal:99,
  vikhlinin_etal:06, zhang_etal:08, vikhlinin_etal:09, arnaud_etal:07,
  arnaud_etal:09, sun_etal:09}). Tight correlations between spherical
overdensity mass and observables are crucial for a robust
interpretation of the observed cluster counts and clustering in
deriving cosmological constraints. The scatter of mass-observable
relations may depend on the value of $\D$. In addition, particular
observations may only extend out to a limited radius corresponding to
$\Delta$ considerably higher than the often used virial value of
$\Delta\approx 200$. Thus, we seek to calibrate a fitting function
that can be adapted to any value of $\Delta$.

\cite{hu_kravtsov:03} and \cite{manera_etal:09} compared existing halo
bias models to SO N-body results at the cluster mass scale. But
previous studies to calibrate halo bias on numerical simulations have
focused exclusively on the friends-of-friends (FOF) halo finding
algorithm (\citealt{jing:98, jing:99, sheth_tormen:99, smt:01,
  seljak_warren:04, tinker_etal:05, pillepich_etal:08,
  reed_etal:08}). The FOF algorithm is a percolation scheme that makes
no assumptions about halo geometry, but may spuriously group distinct
halos together into the same object, confusing the comparison between
cluster observables theoretical results (\citealt{white:01,
  tinker_etal:08_mf, lukic_etal:09}). Additionally, previous
calibrations focus on only one value of the FOF linking length,
$l=0.2$, and thus are not applicable to many mass observables. Galaxy
cluster studies and theoretical halo models benefit from a
self-consistently defined set of coupled mass and bias functions.

The bias of halos is determined by the relative abundance of halos in
different large-scale environments. Thus, theoretical models for halo
bias have been derived from the mass function using the
peak-background split (\citealt{bbks, cole_kaiser:89, mo_white:96,
  sheth_tormen:99, smt:01}). These models produce results that are
reasonably accurate but fail to reproduce in detail the bias of halos
found in numerical simulations (\citealt{jing:98, seljak_warren:04,
  tinker_etal:05, gao_etal:05,
  pillepich_etal:08}). \cite{manera_etal:09} and
\cite{manera_gaztanaga:09} demonstrate that using the peak background
split to calculate the bias of massive halos from their mass function
does not accurately match the clustering as measured from their
spatial distribution. In addition to calibrating the functional form
of the bias, we test the peak background split.

In \S 2 we summarize our list of simulations and the numerical
techniques for calculating bias. In \S 3 we present our fitting
formulae for large scale bias, comparing to previous works and
exploring any redshift evolution. In \S 4 we use our results to test
the peak-background split. In \S 5 we summarize our results.

%%%%%%%%%%%%%%%%%%%%%%%%%%%
% TABLE 1 TABLE 1 TABLE 1
%%%%%%%%%%%%%%%%%%%%%%%%%%%

\begin{deluxetable*}{cccclccclc}
\tablecolumns{7} 
\tablewidth{40pc} 
\tablecaption{Properties of the Simulation Set}
\tablehead{\colhead{$L_{\rm box}$ \hmpc} & \colhead{Name} &  \colhead{$\epsilon$ \hkpc} & \colhead{$N_p$} &\colhead{$m_p$ \hmsol} & \colhead{$(\om,\Omega_b,\sigma_8,h,n)$} & \colhead{Code} & \colhead{$z_i$} & \colhead{$z_{\rm out}$} & \colhead{$\D_{\rm max}$} }

\startdata

768 & H768 & 25 & $1024^3$ &$3.51\times 10^{10}$ & $(0.3,0.04,0.9,0.7,1)$ & HOT& 40 & 0 & 800\\
384 & H384 & 14 & $1024^3$ &$4.39\times 10^{9}$ & $(0.3,0.04,0.9,0.7,1)$ & HOT& 48 & 0 & 3200\\
271 & H271 & 10 & $1024^3$ &$1.54\times 10^{9}$ & $(0.3,0.04,0.9,0.7,1)$ & HOT& 51 & 0 &  3200\\
192 & H192 & 4.9 & $1024^3$ &$5.89\times 10^{8}$ & $(0.3,0.04,0.9,0.7,1)$ & HOT& 54 & 0 &  3200\\
96 & H96 & 1.4 & $1024^3$ &$6.86\times 10^{7}$ & $(0.3,0.04,0.9,0.7,1)$ & HOT& 65 & 0 & 3200\\
\hline
1280 & L1280 & 120 & $640^3$ &$5.99\times 10^{11}$ & $(0.27,0.04,0.9,0.7,1)$ & GADGET2 & 49 & 0, 0.5, 1.0 & 600\\
500 & L500 & 15 & $1024^3\times$2 & $8.24\times 10^{9}$ & $(0.3,0.045,0.9,0.7,1)$ & GADGET2 & 40& 0, 0.5, 1.25, 2.5 & 3200\\
250 & L250 & 7.6 & $512^3$ & $9.69\times 10^{9}$ & $(0.3,0.04,0.9,0.7,1)$ & ART & 49 & 0, 0.5, 1.25, 2.5 & 3200 \\
120 & L120 & 1.8 & $512^3$ & $1.07\times 10^{9}$ & $(0.3,0.04,0.9,0.7,1)$ & ART & 49 & 0, 0.5, 1.25, 2.5 & 3200 \\
80 & L80 & 1.2 & $512^3$ & $3.18\times 10^{8}$ & $(0.3,0.04,0.9,0.7,1)$ & ART & 49 & 0, 0.5, 1.25, 2.5 & 3200\\
\hline
1000 & L1000W & 30 & $1024^3$ & $6.98\times 10^{10}$ & $(0.27,0.044,0.79,0.7,0.95)$ & ART& 60 & 0, 0.5, 1.0, 1.25 & 3200\\
384 & H384W & 14 & $1024^3$ & $3.80\times 10^{9}$ & $(0.26,0.044,0.75,0.71,0.94)$ & HOT& 35 & 0 & 3200 \\
384 & H384$\om$ & 14 & $1024^3$ & $2.92\times 10^{9}$ & $(0.2,0.04,0.9,0.7,1)$& HOT& 42 & 0 &  3200\\
120 & L120W & 0.9 & $1024^3$ & $1.21\times 10^8$ & $(0.27,0.044,0.79,0.7,0.95)$ & ART& 100 & 1.25, 2.5 & 3200\\
80 & L80W & 1.2 & $512^3$ & $2.44\times 10^{8}$ & $(0.23,0.04,0.75,0.73,0.95)$ & ART& 49 & 0, 0.5, 1.25, 2.5 & 3200\\

\enddata \tablecomments{ The top set of 5 simulations are from the
  \cite{warren_etal:06} study. The second list of 5 simulations are of
  the same WMAP1 cosmology, but with different numerical codes. The
  third list of 5 simulations are of alternate cosmologies, focusing
  on the WMAP3 parameter set. The HOT code employs Plummer softening,
  while GADGET employs spline softening. The values of $\epsilon$
  listed for the GADGET simulations are the equivalent Plummer
  softening; when calculating the spline softening kernel, GADGET uses
  a value of 1.4$\epsilon$. The force resolution of the ART code is
  based on the size of the grid cell at the highest level of
  refinement. $\D_{\rm max}$ is the highest overdensity for which halo
  masses can be reliably measured. }
\end{deluxetable*}

\section{Simulations and Methods}

Our simulation set spans a wide range in volume and mass resolution in
order to produce results than span nearly six decades in mass, from
$M\sim 10^{10}$ \hmsol\ halos to massive clusters. The set contains 15
distinct simulations that span local variations of the concordance
\lcdm\ cosmology consistent with results from CMB anisotropies
(\citealt{spergel_etal:03, dunkley_etal:09}). Three numerical codes
are represented in our set; the adaptive refinement technique (ART;
\citealt{kravtsov_etal:97, gottloeber_klypin:08}), the hashed oct-tree
code (HOT; \citealt{warren_etal:06}), and the hybrid tree-particle
mesh code GADGET2 (\citealt{springel:05}). Table 1 lists details of
each simulation, including cosmological parameters, force resolution,
volume, and mass resolution. Further details about the simulation set
can be found in T08. For one simulation, L1280, there are 49
independent realizations. The L1280 simulations were utilized in the
studies on the halo mass function and bias relation of massive halos
in \cite{crocce_etal:06} and \cite{manera_etal:09}, as well as in the
analysis in T08. The dark matter outputs of these simulations were
kindly supplied by R. Scoccimarro.

\cite{crocce_etal:06} point out that improper initial conditions can
result in errors in the resulting halo mass function and, to a lesser
extent, bias function for massive halos. These errors are a product of
starting the simulation at too low a redshift while using first-order
techniques for the initial particle displacements and velocities. The
L1280 simulations utilize second-order perturbation theory to
ameliorate these effects. In T08 we performed multiple re-simulations
of L1000W with initial conditions set using the Zel'dovich
approximation at different redshifts. We found significant differences
between the mass functions measured in the L1000W run with Zel'dovich
initial conditions set at z=30 and the mass function measured in the
L1280 run. However, using Zel'dovich initial conditions at z=60 for
L1000W produces a mass function negligibly different from the L1280
calculations.

Halos are identified in each simulation using the spherical
overdensity (SO) technique outlined in T08. In brief, the code
identifies density peaks in the dark matter and grows spheres around
them until the mean interior density is some set multiple, $\D$, of the
background density. Thus the mass and radius of a halo are related by

\begin{equation}
M_\D=\frac{4}{3}\pi R_\D \bar{\rho}_m(z) \D,
\end{equation}

\noindent where $\bar{\rho}_m(z)$ is the mean density of the universe
at redshift $z$. In our implementation of the SO algorithm, the
spheres that contain halos are allowed to overlap; so long as the
center of one halo is not contained within $R_\D$ of another halo, the
two halos are considered distinct. Owing to small overlaps in the
exteriors of halos, a small fraction of the total mass in halos ($\sim
0.7\%$) is assigned to multiple halos and double counted.  The SO
method of identifying halos makes the halo mass sensitive to the force
resolution of the simulation; if the density profile of a halo is not
properly resolved, the enclosed mass at a given radius will be
smaller. Column 10 in Table 1 lists the maximum value of $\D$ for
which reliable results could be obtained for each simulation. Owing to
the low spatial resolution of the L1280 simulations, our analysis only
utilizes these simulations for $\D=200$. For all simulations, only
halos with more than 400 particles are used. This ensures that all
halos are robustly identified and the halo profiles are well sampled.

We define the bias of dark matter halos as the ratio of the halo power
spectrum to the linear dark matter power spectrum,

\begin{equation}
\label{e.pkbias}
b^2(k) = \frac{P_h(k)}{\plin}.
\end{equation}

\noindent We calculate the power spectrum of each simulation as
follows. The halos of each simulation are binned in a $200^3$ density
mesh using the cloud-in-cell technique, and the power spectrum is
computed through Fourier transformation. All power spectra are
shot-noise subtracted. Aliasing due to the cloud-in-cell grid is
removed through the deconvolution technique outlined in
\cite{jing:05}. Although halo bias is scale-dependent in the
quasi-linear and non-linear regime, here we focus on the large scale
bias, where $b$ is independent of $k$. We calculate $b^2$ as the
average over the 10 largest wavelength modes in the simulation. For
simulations with $L_{\rm box}<200$ \hmpc, non-linearity has set in at
$k\gtrsim 10\times 2\pi/L_{\rm box}$. For these simulations, we
truncate our average to the largest 5 modes. For the $z=2.5$ outputs
of two simulations, L120W and L120, the power spectra do not converge
to a robust asymptotic value within this $k$-range, thus we exclude
these outputs from the analysis. For each simulation, we calculate
$P_h(k)$ for 8 jackknife subsamples of the simulation, removing
density fluctuations from one octant of the box in each subsample. We
use the jackknife subsamples to estimate the error on $b$.

We also check these results against bias as defined by the halo-mass
cross-power spectrum $\bhm=P_{hm}/P_{\rm lin}$. This measure does not
require a shot noise correction, and it yields better statistics when
the halos become vary sparse.

We parameterize our results in terms of `peak height' in the linear
density field, $\nu=\delta_c/\sigma(M)$, where $\delta_c$ is the
critical density for collapse and $\sigma$ is the linear matter
variance on the Lagrangian scale of the halo, ie, $R=(3M/4\pi
\bar{\rho}_m)^{1/3}$, defined as

\begin{equation}
\label{e.sigma}
\sigma^2(R) = \frac{1}{2\pi^2}\int P(k,z)\hat{W}^2(k,R)k^2dk,
\end{equation}

\noindent where $P(k,z)$ is the linear power spectrum at redshift $z$
and $\hat{W}$ is the Fourier transform of the top-hat window function
of radius $R$. In all calculations we use $\delta_c=1.686$. For
reference, $\nu$ of 0.75, 1, 2, and 3 ($\log\nu=[-0.12, 0.0, 0.30,
0.48]$\footnote{Throughout this paper, $\log$ indicates base-10
  logarithm.}) corresponds to $M$ of $2.9\times 10^{11}$, $2.8\times
10^{12}$, $1.2\times 10^{14}$, and $7.0\times 10^{14}$ \hmsol,
respectively, for the L1000W cosmology at $z=0$.

\begin{figure*}
\epsscale{1.0} 
\plotone{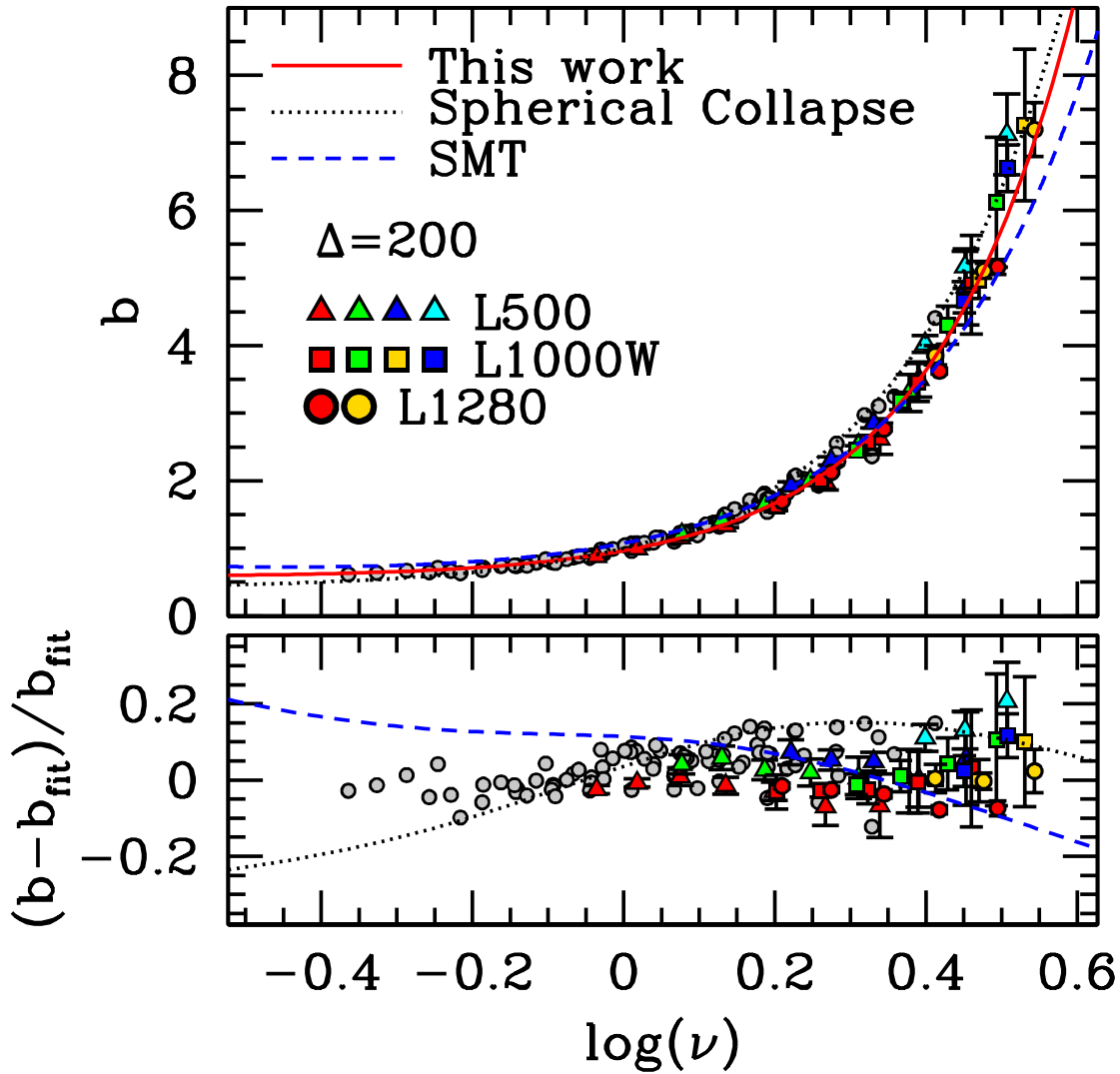}
%\vspace{-1.5cm}
\caption{ \label{sig} {\it Upper Panel:} Large-scale bias as
  determined by the ratio $(P_h/P_{\rm lin})^{1/2}$ for
  $\D=200$. Results from the smaller boxes are represented by the gray
  circles. For these simulations, only measurements with less than
  10\% error are shown to avoid crowding. The larger-volume
  simulations are represented by the colored symbols. Each point type
  indicates a different simulation. The different colors, from left to
  right, go in order of increasing redshift from $z=0$ to $z=2.5$ (see
  Table 1 for the redshift outputs of each simulation). Like colors
  between simulations imply the same redshift. For these large-volume
  simulations, measurements with less than 25\% errors are shown. {\it
    Lower Panel:} Fractional differences of the N-body results with
  the the fitting function shown in the upper panel. }
\end{figure*}

\begin{figure*}
\epsscale{1.0} 
\plotone{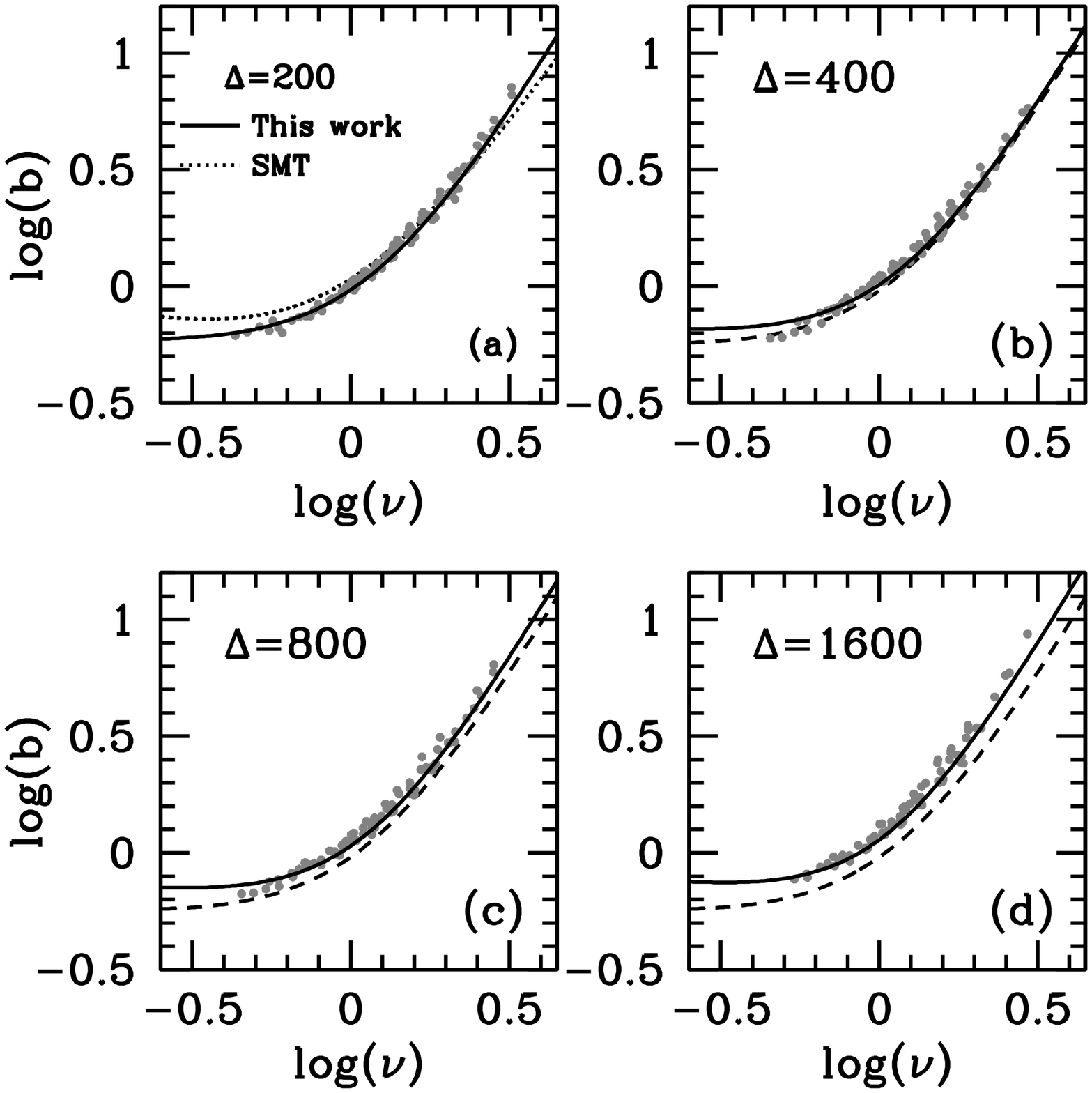}
%\vspace{-1.5cm}
\caption{ \label{bias_delta} Large-scale bias as determined by the
  ratio $(P_h/P_{\rm lin})^{1/2}$ for four values of $\D$. The solid
  line in each panel represents equation (\ref{e.bias}) with the
  $\D$-dependent parameters listed in Table 2. The dotted curve in
  panel (a) is the bias formula of SMT. The dashed curve in panels
  (c)-(d) is the $\D=200$ results (i.e., the solid curve in panel a). }
\end{figure*}

\section{Results}

\subsection{Models and Measurements at $\D=200$}

Figure \ref{sig} shows bias as a function of $\nu$ for all simulations
in Table 1. In this figure, halos are defined with $\D=200$. In the
spherical collapse model, $\D\approx 200$ defines a radius separating
the virialized region and the region of continuing infall in an
$\om=1$ universe (\citealt{lacey_cole:94, eke_etal:98}). This
overdensity is also close to the overdensity of halos identified with
the friends-of-friends (FOF) algorithm with the typical linking
parameter of 0.2 (\citealt{davis_etal:85}). Thus analytic models are
typically compared to numerical results using either $\D=200$ or
FOF(0.2). In Figure \ref{sig}, we compare our $\D=200$ results to two
current models for halo bias from the literature.

First, we compare these results to predictions based on the spherical
collapse model (SC) for the formation of dark matter halos. In SC,
halos collapse when the linear overdensity associated with a peak in
the density field crosses a critical barrier $\delta_c$ independent of
halo mass. \cite{press_schechter:74} used this model to derive an
expression for the mass function of dark matter halos. Using the
peak-background split, which we will describe in more detail in
section \S 4, \cite{cole_kaiser:89} and \cite{mo_white:96} derived a
bias relation of the form

\begin{equation}
\label{e.bias_sc}
b(\nu) = 1 + \frac{\nu^2-1}{\delta_c}.
\end{equation}

However, the Press-Schechter mass function fails to
reproduce the dark matter halo mass function found in simulations
(see, e.g., \citealt{gross_etal:98, lee_shandarin:99, sheth_tormen:99,
  jenkins_etal:01, robertson_etal:09}). Thus it is not surprising that
the bias function in equation (\ref{e.bias_sc}) also does not compare
well to simulations (see, e.g., \citealt{jing:98, jing:99,
  sheth_tormen:99}). In Figure \ref{sig}, the SC model overpredicts
the bias in the range $1\lesssim\nu\lesssim 3$, while underpredicting
slightly the bias for the lowest mass halos in our simulations.

\cite{sheth_tormen:99} (hereafter ST) generalized the expression for
the Press-Schechter mass function and calibrated the free parameters
using numerical simulations. \cite{smt:01} (hereafter SMT) later refined
this calculation, incorporating a ``moving'' barrier for the collapse
criterion of halos in which the critical density varies with the peak
height as motivated by the more physically realistic ellipsoidal
collapse model. Using the peak-background split once again, SMT
derived an improved expression for bias of the form

\begin{eqnarray}
\label{e.smt}
b(\nu) & = & 1 + \frac{1}{\sqrt{a}\dc}\Bigl[\sqrt{a}(a\nu^2)+\sqrt{a}b(a\nu^2)^{1-c} \nonumber\\
& & - \frac{(a\nu^2)^c}{(a\nu^2)^c + b(1-c)(1-c/2)}\Bigr],
\end{eqnarray}

\noindent where $a=0.707$, $b=0.5$, and $c=0.6$. These parameters
describe the shape of the moving barrier. In Figure \ref{sig}, the SMT
bias equation underpredicts the clustering of high-peak halos while
overpredicting the asymptotic bias of low-mass objects. The SMT
function is calibrated using friends-of-friends (FOF) halos, thus the
choice of $\D$ with which to compare is somewhat arbitrary, but it can
be seen that the SMT function and our results will not agree at any
overdensity: SMT bias at low $\nu$ is too high, and is too low at high
$\nu$. When increasing (decreasing) $\D$, the bias at all $\nu$ can
only increase (decrease).

The bias formulae of ST and SMT have been shown before to be
inaccurate at low masses (\citealt{seljak_warren:04, tinker_etal:05,
  gao_etal:05, pillepich_etal:08}). Updated fitting functions have
sometimes used the functional form of ST
(\citealt{mandelbaum_etal:05}) or SMT (\citealt{tinker_etal:05}) with
new parameters chosen to match numerical data, while others have
proposed entirely new functional forms (e.g.,
\citealt{seljak_warren:04, pillepich_etal:08}). Our tests show that
the SMT function does not yield optimal $\chi^2$ values when comparing
to our numerical results. We therefore introduce a similar but more
flexible fitting function of the form

\begin{equation}
\label{e.bias}
b(\nu) = 1 - A\frac{\nu^{a}}{\nu^{a} + \dc^{a}} + B\nu^{b} + C\nu^{c}.
\end{equation}

\noindent Equation (\ref{e.bias}) scales as a power-law at the highest
masses, flattens out at low masses and asymptotes to $b=1$ at $\s=0$,
provided $a>0$.

A convenient property of the SC, ST, and SMT functions is that they
are normalized such that the mean bias of halos is unity. Thus, if one
adopts the halo model ansatz that all mass is contained within halos,
dark matter is not biased against itself. Numerically calibrated bias
functions in the literature do not obey this constraint
(\citealt{jing:98, jing:99, tinker_etal:05, seljak_warren:04,
  pillepich_etal:08}). When fitting for the parameters of equation
(\ref{e.bias}), we enforce this constraint by requiring that our bias
function obey the relation

\begin{equation}
\label{e.normalization}
\int b(\nu)f(\s)d\nu = 1,
\end{equation}

\noindent where $f(\s)$ is the halo mass function, once again
expressed in terms of the scaling variable $\s$. At each $\D$, we use
the halo mass functions listed in Appendix C of T08, which are
normalized such that the mean density of the universe is obtained when
integrating over all halo masses at $z=0$\footnote{The normalized mass
  functions in T08 are expressed in terms of $1/\sigma$ rather than
  $\nu$. For convenience we rewrite this function in terms of $\nu$ in
  Equation (\ref{e.fnu}) and give new mass function parameter values
  in Table 4.}. 

In T08, we found that the mass function is universal at $z=0$ over the
range of cosmologies explored. However, the mass function at higher
redshifts deviates systematically from the $z=0$ results. In Figure
\ref{sig}, we have included the results from {\it all}
redshifts. Although the evolution of $f(\sigma)$ from $z=0$ to $z=1$
is clear in the T08 results, the bias of these halos does not show
significant evolution with redshift. To obtain the parameters of
equation (\ref{e.bias}), we minimize the $\chi^2$ using the jackknife
errors described in the previous section. The best-fit parameters for
the $\D=200$ data, listed in Table 2, yield a $\chi^2$ per degree of
freedom (hereafter $\chinu$) of 1.9 when incorporating all data from
all redshifts. This high value of $\chinu$ is driven by the small
error bars on the L1280 results at $z=0$. With 49 realizations, the
error bars are $\sim 1\%$ at the low-particle limit, thus the few
percent offset between the L1280 results and those of the remaining
simulations yields a high $\chinu$. Removing the L1280 results
(without refitting) yields $\chinu=1.01$. The low spatial resolution
of the L1280 simulations is a possible source of error in the bias
results. Refitting with only the $z=0$ results does not change the
values of the best-fit parameters or change $\chinu$. This implies
that the evolution of bias with redshift is extremely weak between
$0\le z\le 2.5$, if it evolves at all. Simulation to simulation, the
situation is not definitive. The L500 simulations shows an increased
amplitude in the bias of $\sim 5\%$ between $z=0$ and $z=1.25$, but
the L1000W simulation is consistent at all redshifts. Regardless, any
evolution in the bias function at fixed $\nu$ is significantly smaller
than the evolution in the mass function.

\subsection{Large-Scale Bias as a Function of $\D$}

The best-fit parameters of equation (\ref{e.bias}) scale smoothly with
$\D$, allowing us to obtain fitting functions for these parameters as
a function of $\log\D$. The functions that yield the
parameters of equation (\ref{e.bias}) for $200\le \D \le 3200$ are
listed in Table 2. Using these functions, the integral constraint in
equation (\ref{e.normalization}) is satisfied to better than 1\% for
every value of $\D$ considered. If required, higher precision can be
obtained if 5 of the 6 parameters are taken from Table 2 and the last
is solved for numerically.

A comparison between our numerical results and fitting functions for
four values of $\D$ are shown in Figure \ref{bias_delta}. To avoid
crowding and scatter, in each panel we only plot data points with
fractional errors less than 10\%. Figures
\ref{bias_delta}b-\ref{bias_delta}d compare the results for $\D=400$,
800, and 1600 to the $\D=200$ fitting function (shown against the
$\D=200$ data from \ref{bias_delta}a). As $\D$ increases, bias
increases at all mass scales. At high masses this is expected; as $\D$
increases, a fixed set of halos will have lower masses but the same
clustering properties, essentially shifting them along the
$\s$-axis. At low masses, the amplitude of the bias curve also
monotonically increases with $\D$, owing to the substructure within
high mass halos that become distinct objects as $\rd$
decreases. Because these new low mass halos are in the vicinity of
high mass objects, they have significant clustering.\footnote{In
  principle, this makes our results sensitive to the spatial
  resolution of our simulations beyond simply resolving the halo
  density profiles properly. If subhalos are not resolved in some
  subset of our simulations, the change in bias for low $\nu$ halos
  will be underpredicted. Our criterion for including simulations in
  our analysis is that halo density profiles are properly resolved,
  not that substructure is properly resolved. However, the fact that
  bias monotonically increases with $\D$ at low $\nu$ is indicative
  that we are including these ``revealed'' subhalos.}

Table 3 shows the $\chinu$ values for each value of $\D$. The fit is
near $\chinu\approx 1$ at all $\D$, indicating that the fit is
adequate to describing the data even though we have combined {\it all}
the simulation outputs in the fit (i.e., all cosmologies and all
redshifts).

%%%%%%%%%%%%%%%%%%%%%%%%%%%%%%%%%
% Table 2
%%%%%%%%%%%%%%%%%%%%%%%%%%%%%%%%%
\begin{deluxetable}{cc}
  \tablecolumns{2} \tablewidth{10pc} \tablecaption{Parameters of Bias
    Equation (\ref{e.bias}) as a Function of $\D$}
  \tablehead{\colhead{parameter} & \colhead{$f(\D)$} }

\startdata

% as f(sigma)
%A & $1.04 + 0.24x \exp[-(4/x)^4]$ \\
%a & $(x-2.0)0.44$ \\
%B & 0.4 \\
%b & 1.5 \\
%C & $0.07 + 0.4x + 0.7\exp[-(4/x)^4]$ \\
%c & 2.4 \\

% as f(nu)
A & $1.0 + 0.24y \exp[-(4/y)^4]$ \\
a & $0.44y-0.88$\\
B & 0.183 \\
b & 1.5 \\
C & $0.019 + 0.107y + 0.19\exp[-(4/y)^4]$ \\
c & 2.4 \\

\enddata \tablecomments{Note: $y\equiv \log_{10}\D$. }
\end{deluxetable}
%%%%%%%%%%%%%%%%%%%%%%%%%%%%%%%%%
% END Table 2
%%%%%%%%%%%%%%%%%%%%%%%%%%%%%%%%%

\begin{figure*}
\epsscale{1.1}
\plotone{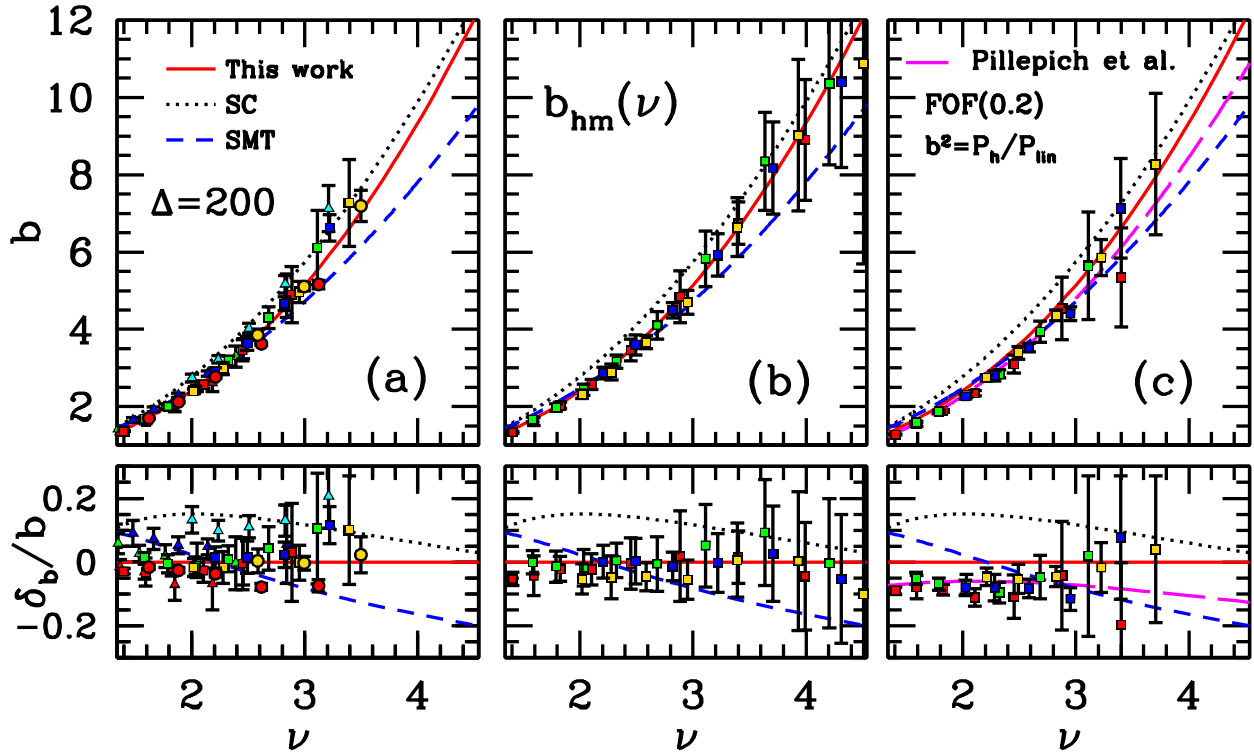}
\vspace{-5.5cm}
\caption{\label{high_bias} {Panel (a)}: The $\D=200$ bias function in
  the high-$\nu$ regime. The points with error bars represent our
  large-volume simulations at the redshifts listed in Table 1. Only
  points with fractional errors less than 25\% are shown. The
  different colors, from left to right, go in order of increasing
  redshift: ({\it red, green, yellow, blue, cyan})$=$(0.0, 0.5, 1.0,
  1.25, 2.5). Like colors between simulations imply the same
  redshift. The dotted line is the spherical collapse prediction. The
  dashed line is the SMT function. The lower panel shows the
  fractional difference with respect to equation (\ref{e.bias}),
  $\delta_b=b_{\rm Nbody}-b_{\rm fit}$. {Panel (b)}: Same as (a), but
  now using bias defined by the ratio of the $P_{hm}/\plin$. Results
  are shown for the L1000W simulation. Colors represent the same
  redshifts as in panel a. Panel (c): Bias of halos identified using
  the FOF algorithm with linking length 0.2. Bias is calculated from
  equation equation (\ref{e.pkbias}). Results are shown for the L1000W
  simulation. Different colors match to different redshifts as
  before. The dotted curve in this Figure is the fitting function of
  \cite{pillepich_etal:08}, which is calibrated on FOF(0.2) halos. }
\end{figure*}

%%%%%%%%%%%%%%%%%%%%%%%%%%%%%%%%%
% Table 3
%%%%%%%%%%%%%%%%%%%%%%%%%%%%%%%%%
\begin{deluxetable}{ccccc}
  \tablecolumns{5} \tablewidth{10pc} \tablecaption{$\chinu$ values of
    the $b(\nu)$ fits} \tablehead{\colhead{} & \colhead{$\D$} &
    \colhead{} & \colhead{$\xdof$} & \colhead{} }

\startdata

& 200 & & 1.01/1.94 & \\
& 300 & & 1.33 &\\
& 400 & & 1.08 &\\
& 600 & & 1.34 &\\
& 800 & & 1.19 &\\
& 1200 & & 1.22& \\
& 1600 & & 1.14 &\\
& 2400 & & 1.06 &\\
& 3200 & & 1.08 &\\

\enddata 
\tablecomments{For $\Delta=200$, the second value of $\chinu$ includes
  the L1280 simulations.}

\end{deluxetable}
%%%%%%%%%%%%%%%%%%%%%%%%%%%%%%%%%
% END Table 3
%%%%%%%%%%%%%%%%%%%%%%%%%%%%%%%%%

%%%%%%%%%%%%%%%%%%%%%%%%%%%%%%%%%
% END Table 4
%%%%%%%%%%%%%%%%%%%%%%%%%%%%%%%%%
\begin{deluxetable*}{rccccc}
\tablecolumns{6} 
\tablewidth{20pc} 
\tablecaption{Parameter of the halo mass function, equation (\ref{e.fnu})}
\tablehead{\colhead{$\D$} & \colhead{$\alpha$} & \colhead{$\beta$} & \colhead{$\gamma$} & \colhead{$\phi$} & \colhead{$\eta$}}

\startdata

200 &  0.368   &   0.589    & 0.864  &   -0.729  &   -0.243 \\
300 & 0.363 &   0.585  &    0.922  &   -0.789  & -0.261 \\
400 & 0.385  &    0.544 &    0.987   &   -0.910   &  -0.261\\
600 & 0.389   &  0.543    &   1.09    & -1.05    & -0.273\\
800 & 0.393  &    0.564   &    1.20  &    -1.20   &  -0.278\\
1200 & 0.365  &    0.623  &    1.34 &   -1.26  &   -0.301\\
1600 &  0.379  &  0.637  &   1.50   &   -1.45  &   -0.301\\
2400 &  0.355  &  0.673 &   1.68  &  -1.50  &   -0.319\\
3200 & 0.327  &   0.702 &   1.81  &  -1.49   &  -0.336\\
\end{deluxetable*}

%%%%%%%%%%%%%%%%%%%%%%%%%%%%%%%%%
% END Table 4
%%%%%%%%%%%%%%%%%%%%%%%%%%%%%%%%%

\subsection{Bias of High-$\nu$ Halos}

The spherical collapse model is defined by a threshold for collapse
that is independent of halo mass. However, peaks in the linear density
field become increasingly elliptical and prolate at low $\nu$,
delaying collapse. Thus, in this mass regime, the barrier in the
ellipsoidal collapse model is significantly higher than the constant
$\delta_c$ assumed in spherical collapse calculations. As a
result, collapsed low-mass halos reside in higher density
environments, making them less abundant and more biased. At high
$\nu$, the ellipsoidal collapse barrier asymptotes to the spherical
$\dc$ value and these two models should thus converge at high
$\nu$. However, the numerically-calibrated barrier used in the SMT fit
asymptotes to a value lower than the spherical collapse $\dc$ in order
to produce the abundance of high-mass halos (see the discussion in
\citealt{robertson_etal:09}). Consequently, the clustering of
high-$\nu$ halos in the SMT model is lower than the spherical collapse
prediction.

In Figure \ref{high_bias}a, we compare our fitting function to the
formulae of the SMT and SC models for halos with $\nu>1.5$. We also
show the bias results from L500 and L1000W for four different
redshifts and from L1280 for two redshifts. We focus on these simulations
because they are the largest in our suite\footnote{We do not include
  the 768 \hmpc\ HOT simulation in this section because the results at
  high $\nu$ are possibly biased due to numerical issues. See the
  discussion in Appendix A of T08. We do include H768 in all fitting,
  but both the mass function and bias relation deviate from the mean
  results at high masses.}. These are the same data presented in
Figure \ref{sig}, but here we are focusing on the high-$\nu$
regime. At $\nu\sim 2$, our simulation results are in good agreement
with the SMT function, but at higher $\nu$, our results rise above the
SMT function and meet the spherical collapse prediction at $\nu\gtrsim
4$. 

At high redshift ($z\sim 10$), \cite{cohn_white:08} found that the
bias of $\nu\sim 3$ halos was better described by the SC models rather
than SMT. However, two other recent studies of halo bias have
concluded in favor of the SMT model for high-peak halos.  In contrast
to \cite{cohn_white:08}, \cite{reed_etal:08} argue that the clustering
of high-$\nu$ halos at $10<z<30$ in their simulations is better
described by the SMT model. They claim that the bias measurements of
\cite{cohn_white:08} are in error because the bias is calculated at
$r=1.5$ \hmpc, where the bias is scale-dependent. To correct for this,
\cite{reed_etal:08} use a fitting function to extrapolate the
translinear correlation function out to linear scales. Using this
technique they find that SMT bias is a better fit than spherical
collapse. \cite{reed_etal:08} are not able to calculate error bars for
their bias values, and the matter variance over the total volume
probed in their simulations is $\sim 12\%$ at the redshifts for which
they obtain their results\footnote{The matter variance of a 1 \hmpc\
  cube at $z=10$ is 39\%, but this is reduced by $\sqrt{11}$ due to
  the 11 realizations they have of this box.}, thus sample variance is
still a concern. The numerical results of \cite{pillepich_etal:08} are
also consistent with SMT at $2\lesssim \nu\lesssim 3$, and deviate
somewhat at higher masses. They use friends-of-friends (FOF) halos
with a linking length of 0.2 times the mean interparticle separation,
and they calculate halo bias by the ratio of the halo-matter power
spectrum, $P_{hm}(k)$, to the matter power
spectrum. \cite{gao_etal:05} and \cite{angulo_etal:08} use the
halo-halo correlation function to determine the bias of FOF(0.2)
halos, with results similar to \cite{pillepich_etal:08}.

In Figure \ref{high_bias}a, our simulations prefer a model that is
intermediate between SC and SMT. But in Figure \ref{high_bias}b we
explore the possible systematics involved in our estimate of
$b(\s)$. Here we use $P_{hm}(k)$ to determine $b(\s)$ from
L1000W. Because shot noise is no longer a concern, this statistic
allows us to extend our bias measurements to higher masses at a given
redshift output. Although the errors at high $\nu$ are large, the
$z=0$ results track our fit at all $\nu$, demonstrating that the these
results are not due to redshift evolution
(and a lack of $z=0$ data at $\nu>2$). The results from other
redshifts are also in agreement with the fit and with the $z=0$
results using $P_{hm}(k)$.

The last systematic to be tested is the choice of halo-finding
algorithm. In Figure \ref{high_bias}c, we plot the bias of halos in
the L1000W simulation that have been identified with the FOF(0.2)
finder. The halos were defined using the same algorithm and linking
length used by both \cite{reed_etal:08} and
\cite{pillepich_etal:08}. Although the difference with $\D=200$ is
small, there is a definite offset between the FOF(0.2) results and our
$\D=200$ fit. At $\nu\sim 3$, the SMT function is a reasonable
description of the data. At higher $\nu$ the numerical results
increase faster than the $\nu^2$ scaling of SMT, but the errors are
too large to see a significant difference with SMT. The empirical fit
determined by \cite{pillepich_etal:08} is also a good fit to our
FOF(0.2) data. Their fit is consistent with SMT at $\nu\sim 2-3$ and
tends higher at larger $\nu$.  As discussed in T08 and
\cite{lukic_etal:09}, a significant fraction of FOF halos are actually
two distinct density peaks linked together by the FOF algorithm. This
linking increases the abundance of massive FOF halos relative to the
abundance of SO halos and reduces the bias. The halo bias of the
FOF(0.168) halo catalogs of \cite{manera_etal:09} agree with the
$\D=200$ results for the L1280 results.

\subsection{NFW Scaling Between Values of $\D$}

\begin{figure}
\epsscale{1.0} 
\plotone{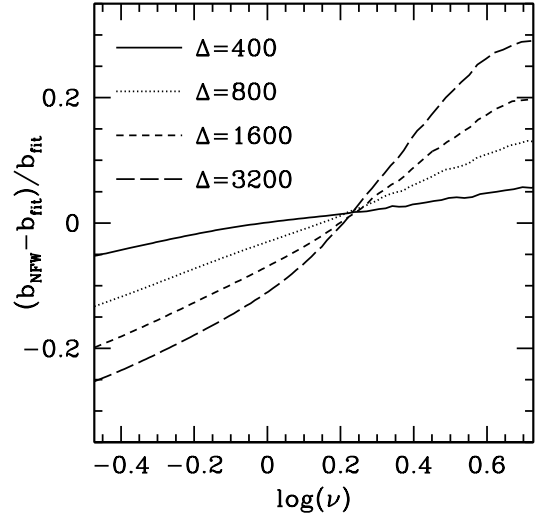}
%\vspace{-1.5cm}
\caption{ \label{bias_scaled} The fractional difference between the
  bias from fitting functions and the bias obtained from rescaling the
  $\D=200$ fitting function to higher overdensities assuming NFW
  profiles and the concentration-mass relation of
  \cite{zhao_etal:09}.}
\end{figure}

One method of obtaining halo statistics at various values of $\D$ is
to assume that halos are described by the density profile of
\cite{nfw:96} (hereafter NFW) and calculate the change in mass between
the desired $\D$ and some fiducial value at which the mass function or
bias relation is calibrated (i.e., \citealt{hu_kravtsov:03}). In T08
we showed that this procedure leads to significant errors in the
inferred mass function at $M<\mstar$, and the abundance of high mass
objects is sensitive to the model for halo concentrations used. In
Figure \ref{bias_scaled} we test this procedure on the bias
function. The curves represent the ratio between the bias obtained
using the fitting functions of Table 2 the bias obtained by taking the
$\D=200$ bias function and rescaling it to higher overdensities. We
assume the concentration-mass relation of \cite{zhao_etal:09} for all
calculations, noting that the results on the high-mass end depend on
the model used. Models that predict a lower concentration for
cluster-sized halos, such as \cite{bullock_etal:01}, yield a much
stronger deviation from the N-body results. Scaling the masses from
one $\D$ to another $\D$ can only result in a horizontal shift of the
bias-mass relation; halos that were substructure at low $\D$ and
revealed at high $\D$ are not taken into account. Low-mass
substuctures in high-mass halos that are ``revealed'' as host halos
when $\D$ increases will increase the mean bias of these objects. This
is why the rescaled bias function underpredicts halo bias at low
$\nu$.

At the high mass end, two effects alter the agreement between the
measured bias and the rescaled bias. For the same object, the
difference in halo mass between two values of $\D$ depends on the
density profile of the halo. Thus, at a given $M_{200}$, scatter in
the concentration-mass relation creates a distribution of halo masses
at higher or lower $\D$. Due to the steepness of the mass function at
$\nu\gtrsim 1$, more low bias halos are scattered up to higher $\nu$
than high bias halos are scattered down. The calculation in Figure
\ref{bias_scaled} assumes only the mean $c(M)$ relation. Scatter
accounts for half of the discrepancy. The rest can be accounted for by
the assembly bias of halos; the effect that halo properties correlate
with large-scale environment (see, e.g., \citealt{sheth_tormen:04,
  gao_etal:05, gao_white:06, wechsler_etal:06}). At $\nu\gtrsim 1$,
more concentrated halos are less clustered than on average of the same
mass. Thus, when scaling halos of the same $M_{200}$ to higher
$M_\Delta$, the value of $M_\Delta$ depends on $c$ and thus depends on
bias such that higher values of $M_\Delta$ are less clustered. Using
the result from \cite{wechsler_etal:06} that a 1-$\sigma$ deviation in
$\log c$ yields a $\sim 13\%$ deviation from the mean clustering for
massive halos, scatter and assembly bias combined bring the rescaled
bias function into agreement with the fitting functions at $\nu\gtrsim
1$.

\section{Testing the Peak-Background Split}

The mass function in Appendix C of T08 is written as a function of
$\sigma$. To match with our parameterization of the bias function in
equation (\ref{e.bias}) and to facilitate the peak-background split,
we rewrite this function in terms of peak height $\nu$. The original
T08 function, $g(\sigma)$, is related to the new function by
$g(\sigma) = \nu f(\nu)$, where

\begin{equation}
\label{e.fnu}
f(\nu) = \alpha\left[1+ \left(\beta\nu\right)^{-2\phi}\right]\nu^{2\eta}e^{-\gamma\nu^2/2}.
\end{equation}

\noindent Table 4 lists the values of the five parameters of equation
(\ref{e.fnu}) for each value of delta.

The mass function parameters in Table 4 are set to match the $z=0$
numerical results from T08. To model the redshift evolution of the
$\D=200$ mass function, the parameters have the following redshift
dependence:

\begin{equation}
\beta = \beta_0\,\left(1+z\right)^{0.20},
\end{equation}

\begin{equation}
\phi = \phi_0\,\left(1+z\right)^{-0.08},
\end{equation}

\begin{equation}
\eta = \eta_0\,\left(1+z\right)^{0.27},
\end{equation}

\begin{equation}
\gamma = \gamma_0\,\left(1+z\right)^{-0.01},
\end{equation}

\noindent where $\beta_0$, etc., is the value of the parameter at
$z=0$ as listed in Table 4. The value of $\alpha$ is obtained through
the integral constraint in equation (\ref{e.normalization}). The
redshift-dependent fitting function is accurate to $\sim 5\%$ at
$\nu>0.6$ relative to the original T08 function. As discussed in T08,
the rate of change of the mass function decreases as $z$ increases,
thus we recommend using the $z=3$ in the above equations to obtain the
mass function at $z>3$.

Theoretical models for the halo mass function assume a one-to-one
correspondence between peaks in the initial density field and
collapsed objects that form at later times. The peak-background split
obtains the bias of halo through the change in the mass function (the
distribution of density peaks) with the large-scale density field (the
background). We implement the peak-background split under the common
assumptions of the excursion set formalism, such as that the smoothing
mass scale (for calculating $\sigma(M)$ in equation [\ref{e.sigma}])
is the same as the mass in the collapsed halo (see
\citealt{zentner:07} for a review).

Following ST, we define the peak height, $\nu_1$,
relative to the background, $\nu_0$, as

\begin{equation}
\label{e.nu_12}
\nu^{2}_{10} \equiv \frac{[\delta_{1} - \delta_{0}]^{2}}{\sigma^{2}_1 - \sigma^{2}_{0}}\approx\nu_{1}^{2}\left(1 - 2\frac{\delta_{0}}{\delta_{1}}\right),
\end{equation}
 
\noindent where on the right hand side we have only kept the leading
order terms. We Taylor expand $\nu_{10}f(\nu_{10})/\nu_1f(\nu_{1})$ to calculate
the Lagrangian halo peak-background split $\deltahL(\nu_1|\delta_0)$.  Using
equation \ref{e.fnu}, the overabundance of halos relative to the mean
in Lagrangian space is

\begin{equation}
  \deltahL(\nu_1|\delta_0) \approx \left[ \gamma \nu_1^{2} - (1 + 2\eta) + \frac{ 2 \phi}{1 + (\beta \nu_1)^{2\phi}}\right]\frac{\delta_{0}}{\delta_{1}}\equiv b_L(\nu_1)\delta_0.
\end{equation} 

\noindent This function is similar to Equation 11 of
\citet{sheth_tormen:99}. The Eulerian bias is related to the Lagrangian
bias by $b_E\equiv 1+b_L$. If we set $\delta_1=\delta_c$, the Eulerian
bias is then

\begin{equation}
\label{e.pb_split}
b(\nu) \approx 1 + \frac{ \gamma \nu^{2} - (1 + 2\eta)}{\deltac} + \frac{ 2 \phi/\deltac}{1 + (\beta \nu)^{2\phi}}.
\end{equation} 

\begin{figure*}
\epsscale{1.0} 
\plotone{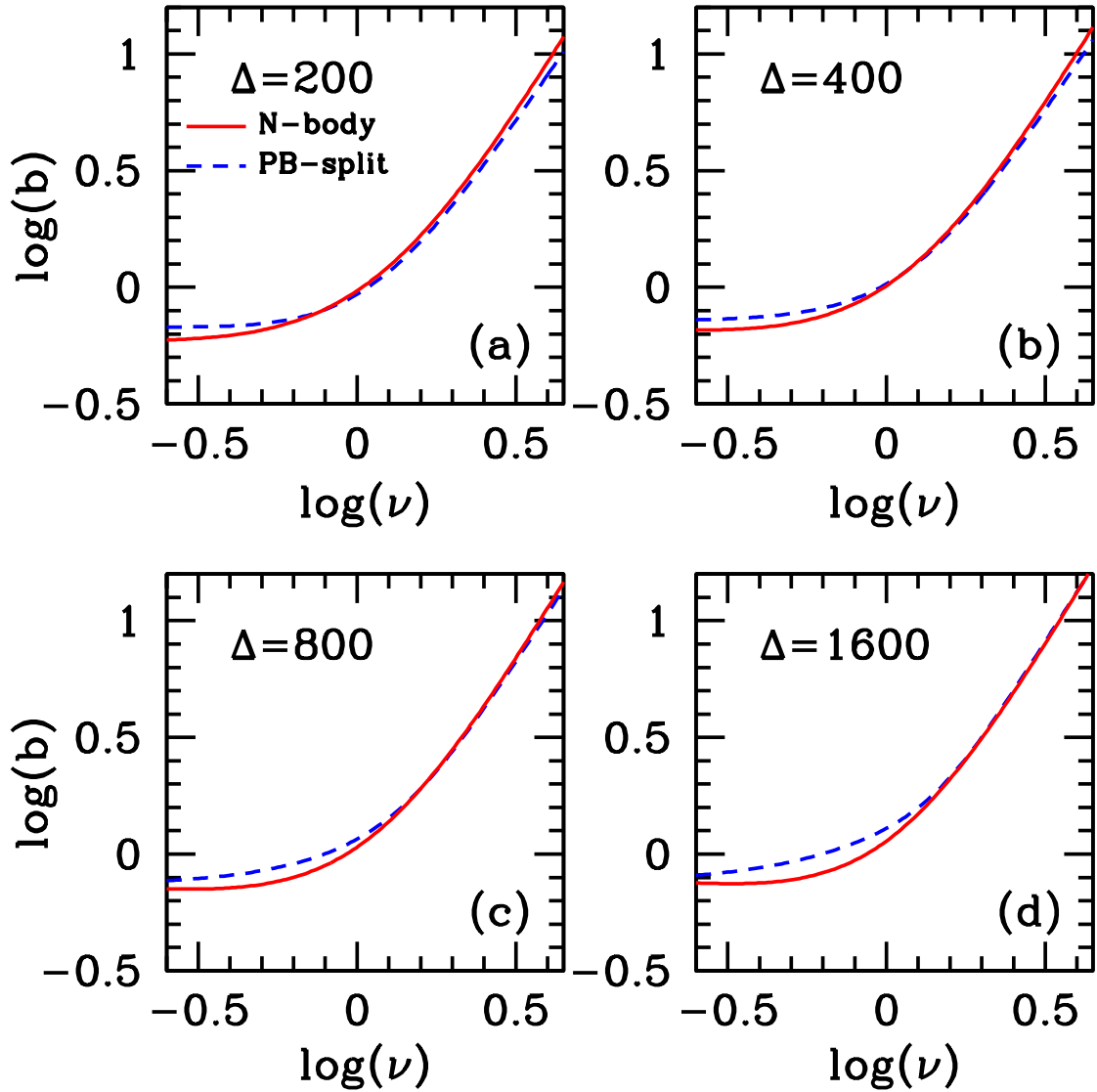}
%\vspace{-1.5cm}
\caption{ \label{pb_split} Comparison of halo bias calibrated from our
  numerical simulations, equation (\ref{e.bias}), with results from
  the peak-background split, equation (\ref{e.pb_split}). At $\D=200$,
  the peak-background split calculation is $\sim 20\%$ high/low and low/high
  $\nu$. As $\D$ increases, the residuals at $\nu>1$ become smaller
  while the residuals at $\nu<1$ become larger. }
\end{figure*}

Figure \ref{pb_split} compares peak-background split bias formula,
equation (\ref{e.pb_split}), to our N-body calibrated results using
equation (\ref{e.bias}). The peak-background split calculation does a
reasonable job modeling the relative change in bias with $\D$; as the
normalization of the mass function is lowered by increasing $\D$, the
amplitude of the bias function over the mass range probed
increases. However, at all overdensities, the peak-background split
overestimates the bias of low-mass halos. For low overdensities,
$\Delta\le 600$, equation (\ref{e.pb_split}) overestimates the bias of
halos above the non-linear mass threshold. For higher overdensities,
the N-body and analytic results appear consistent for high-peak halos,
although the two curves must diverge eventually, as $b\sim \nu^2$ in
the peak-background split and $b \sim\nu^{2.4}$ in our numerical
fit. By definition, equation (\ref{e.pb_split}) satisfies the integral
constraint in equation (\ref{e.normalization}), as does the numerical
fit; at $\log\nu<-0.5$, the bias from equation (\ref{e.bias}) is
higher than the peak-background split calculation.

At high masses at low overdensities, our results are consistent with
those found in \cite{manera_etal:09}. Using FOF-based halos, they find
that employing the peak-background split on the mass function derived
directly from their halo catalogs underpredicts the bias of high-peak
halos. They find this result for three different values of the FOF
linking parameter, 0.2, 0.168, and 0.15. As the linking length is
reduced---which is analogous to increasing the halo overdensity
$\Delta$---the discrepancy is reduced but never completely goes
away. From T08, the linking length best associated with $\Delta=1600$
is 0.1, significantly lower than the three values used by
\cite{manera_etal:09}. Given the trends in their results and in Figure
\ref{pb_split}, we predict that the peak-background split will yield
consistent results at $l\lesssim 0.12$, but only at the high-mass end
of the spectrum.

%%%%%%%%%%%%%%%%%%%%%%%%%%%%%%%%%
% FIGURE
%%%%%%%%%%%%%%%%%%%%%%%%%%%%%%%%%
%%%%%%%%%%%%%%%%%%%%%%%%%%%%%%%%%
% FIGURE
%%%%%%%%%%%%%%%%%%%%%%%%%%%%%%%%%

\section{Summary and Discussion}

We have presented a series of calibrated fitting functions for
large-scale halo bias. The fitting functions are designed to yield the
bias factor for any value of $\D$ within the calibrated range. These
functions are normalized such that, when used in concert with the
normalized mass functions of T08 (given by equations [8]-[12] in this
paper), the overall bias of dark matter is unity. We find a $\sim 6\%$
scatter from simulation-to-simulation. Combined with the $5\%$ error
in the T08 mass function, this level of uncertainty has a
non-negligible impact on the precision with which cosmological
parameters can be constrained from cluster abundance studies; the Dark
Energy Figure-of-Merit (\citealt{detf}) is reduced by 25-50\%,
depending on the details of the survey
(\citealt{cunha_evrard:09,heidi_etal:09}). More importantly, T08 and
this study focus exclusively on cosmological parameters in which the
vacuum energy density is constant with redshift. More study is
required to determine if the halo bias function is universal with
variations in universal expansion and growth history induced by dark
energy. For cluster studies, where the primary concern is the
abundance of massive objects, a series of large-volume simulations
similar to L1000W are required to address this uncertainty. To isolate
the effects of dark energy in both the mass function and bias
function, using the same initial phases with different dark energy
equations of state would eliminate sample variance, which is a concern
even for $h^{-1}$\,Gpc simulations.

Within the precision of our data set, the numerical results do not
show evidence for significant evolution of bias with redshift. Any
evolution must be at the $\lesssim 5\%$ level over our redshift
baseline. This finding contrasts with our results from the mass
function; in T08 we demonstrated that the SO mass function evolves by
up to $\sim 50\%$ from $z=0$ to $z=2.5$. This evolution is more
pronounced with higher overdensity. If the abundance of dark matter
halos is connected to the bias of halos---as is assumed in the
peak-background split---one would assume that $b$ should increase at
fixed $\nu$ as redshift increases. To the statistical precision of our
data, however, halo bias can be modeled by a single,
redshift-independent function.

Although the absolute predictions of the peak-background split fail to
reproduce our numerical results in detail, this method reasonably
tracks the change in the bias function with $\D$. Thus we can gain
insight from using the peak-background split on the mass function at
various redshifts to see how it changes under the peak-background
ansatz. In T08, the evolution in the mass function is mostly
encompassed by a change in the overall normalization of $\nu f(\nu)$
(cf., Figure 6 in T08), with a slightly stronger evolution for
$\nu\gtrsim 1$ halos. A change in the overall abundance of halos does
not induce a change in their clustering. Thus, employing the
peak-background split on the redshift-dependent mass function for
$\D=200$ at $z=1.25$ yields a bias function that is at nearly
identical to the $z=0$ peak-background split function at high $\nu$,
and is only $\sim 5\%$ higher at $\nu\lesssim 0.4$.

We have paid significant attention to the bias of halos at $\nu\gtrsim
2$, which corresponds to the peak-height for galaxy clusters. Our
$\D=200$ halo catalogs disfavor a bias function with an amplitude as
low as SMT. This result is robust to any choice of statistic with
which to calculate the bias. The numerical results of
\cite{reed_etal:08} and \cite{pillepich_etal:08} find good agreement
with SMT at these scales, but these results are based on FOF(0.2)
halos. The known problem of linking distinct objects in the FOF
algorithm would reduce bias at fixed mass because two (or more)
objects with intrinsically lower bias are being counted as one more
massive object. In addition, as pointed out in \cite{lukic_etal:09},
the mean ratio between FOF halo mass and SO halo mass depends on
concentration even for unbridged samples of halos, so this will also
affect the relative bias between the two mass definitions in a
non-trivial manner.  In our simulations, $\D=200$ and FOF(0.2) do not
agree. At $\nu=3$, our FOF(0.2) results appear to be in agreement with
the SMT function as well as the fitting function of
\cite{pillepich_etal:08}.

In a general sense, the peak-background split does achieve marked
success; the first-order derivation calculated here is accurate to
$\lesssim 20\%$ and correctly predicts the change in bias with
$\Delta$. There are several possibilities in explaining the
differences between the theory and N-body results. For massive halos,
a first-order expansion of the peak-background split may not be
sufficient. However, \cite{manera_etal:09} and
\cite{manera_gaztanaga:09} demonstrate that higher-order terms do
increase the accuracy of the calculation at high masses, but decreases
it at lower masses. The growth of low-mass halos in overdensities is
truncated due to the presence of nearby, high-mass objects
(\citealt{wechsler_etal:06, wang_mo_jing:07, dalal_etal:08,
  hahn_etal:09}). Our implementation of the peak-background split
assumes that the local peak corresponding to the collapse threshold is
$\delta_1=\delta_c=1.686$, ignoring any environmental effects on the
collapse of dark matter halos. Alternatively, as discussed in
\cite{manera_etal:09}, it is not clear that the mass that enters into
the calculation of the peak height, $\delta_c/\sigma(M)$, should be
the same mass of the object that eventually collapses. The mass
contained within the peak does not completely map onto the mass within
the collapsed halo (\citealt{dalal_etal:08}). It remains to be seen
whether a more robust implementation of the peak-background split
model, in which the Taylor expansion is replaced with a more rigorous
treatment, can reconcile the differences between theory and numerical
results, or if the peak-background split fails at a more fundamental
level. More work is required to isolate the failures of the model and
bring our theoretical understanding of the formation of dark matter
halos into agreement with the ever-increasing precision of numerical
simulations.

\acknowledgements We thank Roman Scoccimarro for sharing his N-body
simulations and for the computational resources to analyze them. BER
is supported by a Hubble Fellowship grant, program number
HST-HF-51262.01-A provided by NASA from the Space Telescope Science
Institute, which is operated by the Association of Universities for
Research in Astronomy, Incorporated, under NASA contract NAS5-26555.
AVK is supported by the NSF under grants No.  AST-0239759 and
AST-0507666, by NASA through grant NAG5-13274, and by the Kavli
Institute for Cosmological Physics at the University of Chicago.
Portions of this work were performed under the auspices of the
U.S. Dept. of Energy, and supported by its contract \#W-7405-ENG-36 to
Los Alamos National Laboratory.  Computational resources were provided
by the LANL open supercomputing initiative. SG acknowledges support
by the German Academic Exchange Service.  Some of the simulations were
performed at the Leibniz Rechenzentrum Munich, partly using German
Grid infrastructure provided by AstroGrid-D. The GADGET SPH
simulations have been done in the MareNostrum supercomputer at BSC-CNS
(Spain) and analyzed at NIC J\"ulich (Germany).  G.Y. and S.G. wish to
thank A.I. Hispano-Alemanas and DFG for financial
support. G.Y. acknowledges support also from M.E.C. grants
FPA2006-01105 and AYA2006-15492-C03.

%%%%%%%%%%%%%%%%%%%%%%%%%%%%%%%%%%%%%%%%%%%%%%%%%%%%%%%%%%%%%%%%%%%%%%%%
%  Bibliography
%%%%%%%%%%%%%%%%%%%%%%%%%%%%%%%%%%%%%%%%%%%%%%%%%%%%%%%%%%%%%%%%%%%%%%%%

\pagebreak
\bibliography{../risa}

\begin{thebibliography}{71}
\expandafter\ifx\csname natexlab\endcsname\relax\def\natexlab#1{#1}\fi

\bibitem[{{Abazajian} {et~al.}(2005){Abazajian}, {Zheng}, {Zehavi}, {Weinberg},
  {Frieman}, {Berlind}, {Blanton}, {Bahcall}, {Brinkmann}, {Schneider}, \&
  {Tegmark}}]{kev_etal:05}
{Abazajian}, K., {Zheng}, Z., {Zehavi}, I., {Weinberg}, D.~H., {Frieman},
  J.~A., {Berlind}, A.~A., {Blanton}, M.~R., {Bahcall}, N.~A., {Brinkmann}, J.,
  {Schneider}, D.~P., \& {Tegmark}, M. 2005, \apj, 625, 613

\bibitem[{{Albrecht} {et~al.}(2006){Albrecht}, {Bernstein}, {Cahn}, {Freedman},
  {Hewitt}, {Hu}, {Huth}, {Kamionkowski}, {Kolb}, {Knox}, {Mather}, {Staggs},
  \& {Suntzeff}}]{detf}
{Albrecht}, A., {Bernstein}, G., {Cahn}, R., {Freedman}, W.~L., {Hewitt}, J.,
  {Hu}, W., {Huth}, J., {Kamionkowski}, M., {Kolb}, E.~W., {Knox}, L.,
  {Mather}, J.~C., {Staggs}, S., \& {Suntzeff}, N.~B. 2006,
  arXiv:astro-ph/0609591

\bibitem[{{Angulo} {et~al.}(2008){Angulo}, {Baugh}, \&
  {Lacey}}]{angulo_etal:08}
{Angulo}, R.~E., {Baugh}, C.~M., \& {Lacey}, C.~G. 2008, \mnras, 387, 921

\bibitem[{{Arnaud} {et~al.}(2007){Arnaud}, {Pointecouteau}, \&
  {Pratt}}]{arnaud_etal:07}
{Arnaud}, M., {Pointecouteau}, E., \& {Pratt}, G.~W. 2007, \aap, 474, L37

\bibitem[{{Arnaud} {et~al.}(2009){Arnaud}, {Pratt}, {Piffaretti}, {Boehringer},
  {Croston}, \& {Pointecouteau}}]{arnaud_etal:09}
{Arnaud}, M., {Pratt}, G.~W., {Piffaretti}, R., {Boehringer}, H., {Croston},
  J.~H., \& {Pointecouteau}, E. 2009, \aa, (submitted), ArXiv:0910.1234

\bibitem[{{Bardeen} {et~al.}(1986){Bardeen}, {Bond}, {Kaiser}, \&
  {Szalay}}]{bbks}
{Bardeen}, J.~M., {Bond}, J.~R., {Kaiser}, N., \& {Szalay}, A.~S. 1986, \apj,
  304, 15

\bibitem[{{Bialek} {et~al.}(2001){Bialek}, {Evrard}, \&
  {Mohr}}]{bialek_etal:01}
{Bialek}, J.~J., {Evrard}, A.~E., \& {Mohr}, J.~J. 2001, \apj, 555, 597

\bibitem[{{Bullock} {et~al.}(2001){Bullock}, {Kolatt}, {Sigad}, {Somerville},
  {Kravtsov}, {Klypin}, {Primack}, \& {Dekel}}]{bullock_etal:01}
{Bullock}, J.~S., {Kolatt}, T.~S., {Sigad}, Y., {Somerville}, R.~S.,
  {Kravtsov}, A.~V., {Klypin}, A.~A., {Primack}, J.~R., \& {Dekel}, A. 2001,
  \mnras, 321, 559

\bibitem[{{Cohn} \& {White}(2008)}]{cohn_white:08}
{Cohn}, J.~D. \& {White}, M. 2008, \mnras, 385, 2025

\bibitem[{{Cole} \& {Kaiser}(1989)}]{cole_kaiser:89}
{Cole}, S. \& {Kaiser}, N. 1989, \mnras, 237, 1127

\bibitem[{{Crocce} {et~al.}(2006){Crocce}, {Pueblas}, \&
  {Scoccimarro}}]{crocce_etal:06}
{Crocce}, M., {Pueblas}, S., \& {Scoccimarro}, R. 2006, \mnras, 373, 369

\bibitem[{{Cunha} \& {Evrard}(2009)}]{cunha_evrard:09}
{Cunha}, C.~E. \& {Evrard}, A.~E. 2009, \prd, submitted, ArXiv:0908.0526

\bibitem[{{da Silva} {et~al.}(2004){da Silva}, {Kay}, {Liddle}, \&
  {Thomas}}]{da_silva_etal:04}
{da Silva}, A.~C., {Kay}, S.~T., {Liddle}, A.~R., \& {Thomas}, P.~A. 2004,
  \mnras, 348, 1401

\bibitem[{{Dalal} {et~al.}(2008){Dalal}, {White}, {Bond}, \&
  {Shirokov}}]{dalal_etal:08}
{Dalal}, N., {White}, M., {Bond}, J.~R., \& {Shirokov}, A. 2008, \apj, 687, 12

\bibitem[{{Davis} {et~al.}(1985){Davis}, {Efstathiou}, {Frenk}, \&
  {White}}]{davis_etal:85}
{Davis}, M., {Efstathiou}, G., {Frenk}, C.~S., \& {White}, S. D.~M. 1985, \apj,
  292, 371

\bibitem[{{Dunkley} {et~al.}(2009){Dunkley}, {Komatsu}, {Nolta}, {Spergel},
  {Larson}, {Hinshaw}, {Page}, {Bennett}, {Gold}, {Jarosik}, {Weiland},
  {Halpern}, {Hill}, {Kogut}, {Limon}, {Meyer}, {Tucker}, {Wollack}, \&
  {Wright}}]{dunkley_etal:09}
{Dunkley}, J., {Komatsu}, E., {Nolta}, M.~R., {Spergel}, D.~N., {Larson}, D.,
  {Hinshaw}, G., {Page}, L., {Bennett}, C.~L., {Gold}, B., {Jarosik}, N.,
  {Weiland}, J.~L., {Halpern}, M., {Hill}, R.~S., {Kogut}, A., {Limon}, M.,
  {Meyer}, S.~S., {Tucker}, G.~S., {Wollack}, E., \& {Wright}, E.~L. 2009,
  \apjs, 180, 306

\bibitem[{{Eke} {et~al.}(1998){Eke}, {Cole}, {Frenk}, \& {Patrick
  Henry}}]{eke_etal:98}
{Eke}, V.~R., {Cole}, S., {Frenk}, C.~S., \& {Patrick Henry}, J. 1998, \mnras,
  298, 1145

\bibitem[{{Gao} {et~al.}(2005){Gao}, {Springel}, \& {White}}]{gao_etal:05}
{Gao}, L., {Springel}, V., \& {White}, S.~D.~M. 2005, MNRAS, 363, L66

\bibitem[{{Gao} \& White(2006)}]{gao_white:06}
{Gao}, L. \& White, S. D.~M. 2006

\bibitem[{{Gottl\"ober} \& {Klypin}(2008)}]{gottloeber_klypin:08}
{Gottl\"ober}, S. \& {Klypin}, A. 2008, ArXiv:0803.4343

\bibitem[{{Gross} {et~al.}(1998){Gross}, {Somerville}, {Primack}, {Holtzman},
  \& {Klypin}}]{gross_etal:98}
{Gross}, M.~A.~K., {Somerville}, R.~S., {Primack}, J.~R., {Holtzman}, J., \&
  {Klypin}, A. 1998, \mnras, 301, 81

\bibitem[{{Hahn} {et~al.}(2009){Hahn}, {Porciani}, {Dekel}, \&
  {Carollo}}]{hahn_etal:09}
{Hahn}, O., {Porciani}, C., {Dekel}, A., \& {Carollo}, C.~M. 2009, \mnras, 398,
  1742

\bibitem[{{Hu} \& {Kravtsov}(2003)}]{hu_kravtsov:03}
{Hu}, W. \& {Kravtsov}, A.~V. 2003, \apj, 584, 702

\bibitem[{{Jenkins} {et~al.}(2001){Jenkins}, {Frenk}, {White}, {Colberg},
  {Cole}, {Evrard}, {Couchman}, \& {Yoshida}}]{jenkins_etal:01}
{Jenkins}, A., {Frenk}, C.~S., {White}, S. D.~M., {Colberg}, J.~M., {Cole}, S.,
  {Evrard}, A.~E., {Couchman}, H. M.~P., \& {Yoshida}, N. 2001, \mnras, 321,
  372

\bibitem[{{Jing}(1998)}]{jing:98}
{Jing}, Y.~P. 1998, \apjl, 503, L9+

\bibitem[{{Jing}(1999)}]{jing:99}
---. 1999, \apjl, 515, L45

\bibitem[{{Jing}(2005)}]{jing:05}
---. 2005, \apj, 620, 559

\bibitem[{{Kaiser}(1984)}]{kaiser:84}
{Kaiser}, N. 1984, \apjl, 284, L9

\bibitem[{{Kravtsov} {et~al.}(1997){Kravtsov}, {Klypin}, \&
  {Khokhlov}}]{kravtsov_etal:97}
{Kravtsov}, A.~V., {Klypin}, A.~A., \& {Khokhlov}, A.~M. 1997, ApJS, 111, 73

\bibitem[{{Kravtsov} {et~al.}(2006){Kravtsov}, {Vikhlinin}, \&
  {Nagai}}]{kravtsov_etal:06}
{Kravtsov}, A.~V., {Vikhlinin}, A., \& {Nagai}, D. 2006, \apj, 650, 128

\bibitem[{{Lacey} \& {Cole}(1994)}]{lacey_cole:94}
{Lacey}, C. \& {Cole}, S. 1994, \mnras, 271, 676

\bibitem[{{Lee} \& {Shandarin}(1999)}]{lee_shandarin:99}
{Lee}, J. \& {Shandarin}, S.~F. 1999, \apjl, 517, L5

\bibitem[{{Lima} \& {Hu}(2004)}]{lima_hu:04}
{Lima}, M. \& {Hu}, W. 2004, \prd, 70, 043504

\bibitem[{{Lima} \& {Hu}(2005)}]{lima_hu:05}
---. 2005, \prd, 72, 043006

\bibitem[{{Luki{\'c}} {et~al.}(2009){Luki{\'c}}, {Reed}, {Habib}, \&
  {Heitmann}}]{lukic_etal:09}
{Luki{\'c}}, Z., {Reed}, D., {Habib}, S., \& {Heitmann}, K. 2009, \apj, 692,
  217

\bibitem[{{Majumdar} \& {Mohr}(2004)}]{majumdar_mohr:04}
{Majumdar}, S. \& {Mohr}, J.~J. 2004, \apj, 613, 41

\bibitem[{{Mandelbaum} {et~al.}(2005){Mandelbaum}, {Tasitsiomi}, {Seljak},
  {Kravtsov}, \& {Wechsler}}]{mandelbaum_etal:05}
{Mandelbaum}, R., {Tasitsiomi}, A., {Seljak}, U., {Kravtsov}, A.~V., \&
  {Wechsler}, R.~H. 2005, \mnras, 362, 1451

\bibitem[{{Manera} \& {Gaztanaga}(2009)}]{manera_gaztanaga:09}
{Manera}, M. \& {Gaztanaga}, E. 2009, \mnras, submitted, ArXiv:0912.0446

\bibitem[{{Manera} {et~al.}(2009){Manera}, {Sheth}, \&
  {Scoccimarro}}]{manera_etal:09}
{Manera}, M., {Sheth}, R.~K., \& {Scoccimarro}, R. 2009, \mnras (in press),
  arXiv:0906.1314, 1826

\bibitem[{{Mo} \& {White}(1996)}]{mo_white:96}
{Mo}, H.~J. \& {White}, S.~D.~M. 1996, \mnras, 282, 347

\bibitem[{{Mohr} {et~al.}(1999){Mohr}, {Mathiesen}, \& {Evrard}}]{mohr_etal:99}
{Mohr}, J.~J., {Mathiesen}, B., \& {Evrard}, A.~E. 1999, \apj, 517, 627

\bibitem[{{Nagai}(2006)}]{nagai:06}
{Nagai}, D. 2006, \apj, 650, 538

\bibitem[{{Navarro} {et~al.}(1996){Navarro}, {Frenk}, \& {White}}]{nfw:96}
{Navarro}, J.~F., {Frenk}, C.~S., \& {White}, S. D.~M. 1996, \apj, 462, 563

\bibitem[{{Oguri}(2009)}]{oguri:09}
{Oguri}, M. 2009, Physical Review Letters, 102, 211301

\bibitem[{{Pillepich} {et~al.}(2008){Pillepich}, {Porciani}, \&
  {Hahn}}]{pillepich_etal:08}
{Pillepich}, A., {Porciani}, C., \& {Hahn}, O. 2008, ArXiv:0811.4176

\bibitem[{{Press} \& {Schechter}(1974)}]{press_schechter:74}
{Press}, W.~H. \& {Schechter}, P. 1974, \apj, 187, 425

\bibitem[{{Reed} {et~al.}(2008){Reed}, {Bower}, {Frenk}, {Jenkins}, \&
  {Theuns}}]{reed_etal:08}
{Reed}, D.~S., {Bower}, R., {Frenk}, C.~S., {Jenkins}, A., \& {Theuns}, T.
  2008, ArXiv e-prints, arXiv:0804.0004 [astro-ph]

\bibitem[{{Robertson} {et~al.}(2009){Robertson}, {Kravtsov}, {Tinker}, \&
  {Zentner}}]{robertson_etal:09}
{Robertson}, B.~E., {Kravtsov}, A.~V., {Tinker}, J., \& {Zentner}, A.~R. 2009,
  \apj, 696, 636

\bibitem[{{Seljak} \& {Warren}(2004)}]{seljak_warren:04}
{Seljak}, U. \& {Warren}, M.~S. 2004, \mnras, 355, 129

\bibitem[{{Sheth} {et~al.}(2001){Sheth}, {Mo}, \& {Tormen}}]{smt:01}
{Sheth}, R.~K., {Mo}, H.~J., \& {Tormen}, G. 2001, \mnras, 323, 1

\bibitem[{{Sheth} \& {Tormen}(1999)}]{sheth_tormen:99}
{Sheth}, R.~K. \& {Tormen}, G. 1999, \mnras, 308, 119

\bibitem[{{Sheth} \& {Tormen}(2004)}]{sheth_tormen:04}
---. 2004, MNRAS, 350, 1385

\bibitem[{{Spergel} {et~al.}(2003){Spergel}, {Verde}, {Peiris}, {Komatsu},
  {Nolta}, {Bennett}, {Halpern}, {Hinshaw}, {Jarosik}, {Kogut}, {Limon},
  {Meyer}, {Page}, {Tucker}, {Weiland}, {Wollack}, \&
  {Wright}}]{spergel_etal:03}
{Spergel}, D.~N., {Verde}, L., {Peiris}, H.~V., {Komatsu}, E., {Nolta}, M.~R.,
  {Bennett}, C.~L., {Halpern}, M., {Hinshaw}, G., {Jarosik}, N., {Kogut}, A.,
  {Limon}, M., {Meyer}, S.~S., {Page}, L., {Tucker}, G.~S., {Weiland}, J.~L.,
  {Wollack}, E., \& {Wright}, E.~L. 2003, \apjs, 148, 175

\bibitem[{{Springel}(2005)}]{springel:05}
{Springel}, V. 2005, \mnras, 364, 1105

\bibitem[{{Sun} {et~al.}(2009){Sun}, {Voit}, {Donahue}, {Jones}, {Forman}, \&
  {Vikhlinin}}]{sun_etal:09}
{Sun}, M., {Voit}, G.~M., {Donahue}, M., {Jones}, C., {Forman}, W., \&
  {Vikhlinin}, A. 2009, \apj, 693, 1142

\bibitem[{{Tinker} {et~al.}(2008){Tinker}, {Kravtsov}, {Klypin}, {Abazajian},
  {Warren}, {Yepes}, {Gottl{\"o}ber}, \& {Holz}}]{tinker_etal:08_mf}
{Tinker}, J., {Kravtsov}, A.~V., {Klypin}, A., {Abazajian}, K., {Warren}, M.,
  {Yepes}, G., {Gottl{\"o}ber}, S., \& {Holz}, D.~E. 2008, \apj, 688, 709

\bibitem[{{Tinker} {et~al.}(2005){Tinker}, {Weinberg}, {Zheng}, \&
  {Zehavi}}]{tinker_etal:05}
{Tinker}, J.~L., {Weinberg}, D.~H., {Zheng}, Z., \& {Zehavi}, I. 2005, \apj,
  631, 41

\bibitem[{{van den Bosch} {et~al.}(2003){van den Bosch}, {Mo}, \&
  {Yang}}]{vdb_etal:03}
{van den Bosch}, F.~C., {Mo}, H.~J., \& {Yang}, X. 2003, \mnras, 345, 923

\bibitem[{{Vikhlinin} {et~al.}(2006){Vikhlinin}, {Kravtsov}, {Forman}, {Jones},
  {Markevitch}, {Murray}, \& {Van Speybroeck}}]{vikhlinin_etal:06}
{Vikhlinin}, A., {Kravtsov}, A., {Forman}, W., {Jones}, C., {Markevitch}, M.,
  {Murray}, S.~S., \& {Van Speybroeck}, L. 2006, \apj, 640, 691

\bibitem[{{Vikhlinin} {et~al.}(2009){Vikhlinin}, {Kravtsov}, {Burenin},
  {Ebeling}, {Forman}, {Hornstrup}, {Jones}, {Murray}, {Nagai}, {Quintana}, \&
  {Voevodkin}}]{vikhlinin_etal:09}
{Vikhlinin}, A., {Kravtsov}, A.~V., {Burenin}, R.~A., {Ebeling}, H., {Forman},
  W.~R., {Hornstrup}, A., {Jones}, C., {Murray}, S.~S., {Nagai}, D.,
  {Quintana}, H., \& {Voevodkin}, A. 2009, \apj, 692, 1060

\bibitem[{{Wang} {et~al.}(2007){Wang}, {Mo}, \& {Jing}}]{wang_mo_jing:07}
{Wang}, H.~Y., {Mo}, H.~J., \& {Jing}, Y.~P. 2007, \mnras, 375, 633

\bibitem[{{Warren} {et~al.}(2006){Warren}, {Abazajian}, {Holz}, \&
  {Teodoro}}]{warren_etal:06}
{Warren}, M.~S., {Abazajian}, K., {Holz}, D.~E., \& {Teodoro}, L. 2006, \apj,
  646, 881

\bibitem[{{Wechsler} {et~al.}(2006){Wechsler}, {Zentner}, {Bullock},
  {Kravtsov}, \& {Allgood}}]{wechsler_etal:06}
{Wechsler}, R.~H., {Zentner}, A.~R., {Bullock}, J.~S., {Kravtsov}, A.~V., \&
  {Allgood}, B. 2006, \apj, 652, 71

\bibitem[{{White}(2001)}]{white:01}
{White}, M. 2001, \aap, 367, 27

\bibitem[{{Wu} {et~al.}(2009){Wu}, {Zentner}, \& {Wechsler}}]{heidi_etal:09}
{Wu}, H., {Zentner}, A.~R., \& {Wechsler}, R.~H. 2009, \apj, (submitted),
  ArXiv:0910.3668

\bibitem[{{Yoo} {et~al.}(2009){Yoo}, {Weinberg}, {Tinker}, {Zheng}, \&
  {Warren}}]{yoo_etal:09}
{Yoo}, J., {Weinberg}, D.~H., {Tinker}, J.~L., {Zheng}, Z., \& {Warren}, M.~S.
  2009, \apj, 698, 967

\bibitem[{{Zehavi} {et~al.}(2005){Zehavi}, {Zheng}, {Weinberg}, {Frieman},
  {Berlind}, {Blanton}, {Scoccimarro}, {Sheth}, {Strauss}, {Kayo}, {Suto},
  {Fukugita}, {Nakamura}, {Bahcall}, {Brinkmann}, {Gunn}, {Hennessy},
  {Ivezi{\'c}}, {Knapp}, {Loveday}, {Meiksin}, {Schlegel}, {Schneider},
  {Szapudi}, {Tegmark}, {Vogeley}, \& {York}}]{zehavi_etal:05}
{Zehavi}, I., {Zheng}, Z., {Weinberg}, D.~H., {Frieman}, J.~A., {Berlind},
  A.~A., {Blanton}, M.~R., {Scoccimarro}, R., {Sheth}, R.~K., {Strauss}, M.~A.,
  {Kayo}, I., {Suto}, Y., {Fukugita}, M., {Nakamura}, O., {Bahcall}, N.~A.,
  {Brinkmann}, J., {Gunn}, J.~E., {Hennessy}, G.~S., {Ivezi{\'c}}, {\v Z}.,
  {Knapp}, G.~R., {Loveday}, J., {Meiksin}, A., {Schlegel}, D.~J., {Schneider},
  D.~P., {Szapudi}, I., {Tegmark}, M., {Vogeley}, M.~S., \& {York}, D.~G. 2005,
  \apj, 630, 1

\bibitem[{{Zentner}(2007)}]{zentner:07}
{Zentner}, A.~R. 2007, International Journal of Modern Physics D, 16, 763

\bibitem[{{Zhang} {et~al.}(2008){Zhang}, {Finoguenov}, {B{\"o}hringer},
  {Kneib}, {Smith}, {Kneissl}, {Okabe}, \& {Dahle}}]{zhang_etal:08}
{Zhang}, Y.-Y., {Finoguenov}, A., {B{\"o}hringer}, H., {Kneib}, J.-P., {Smith},
  G.~P., {Kneissl}, R., {Okabe}, N., \& {Dahle}, H. 2008, \aap, 482, 451

\bibitem[{{Zhao} {et~al.}(2009){Zhao}, {Jing}, {Mo}, \&
  {B{\"o}rner}}]{zhao_etal:09}
{Zhao}, D.~H., {Jing}, Y.~P., {Mo}, H.~J., \& {B{\"o}rner}, G. 2009, \apj, 707,
  354

\bibitem[{{Zheng} \& {Weinberg}(2007)}]{zheng_weinberg:07}
{Zheng}, Z. \& {Weinberg}, D.~H. 2007, \apj, 659, 1

\end{thebibliography}

\end{document}